\documentclass[fleqn,usenatbib]{mnras}

\usepackage{newtxtext,newtxmath}
\usepackage[T1]{fontenc}

\DeclareRobustCommand{\VAN}[3]{#2}
\let\VANthebibliography\thebibliography
\def\thebibliography{\DeclareRobustCommand{\VAN}[3]{##3}\VANthebibliography}

\usepackage{graphicx}    
\usepackage{tabulary}
\usepackage{footnote}
\usepackage{hyperref}
\usepackage{natbib}
\usepackage[dvipsnames]{xcolor}
\bibpunct{(}{)}{;}{a}{}{,}

\newcommand{\B}[0]{\mathsf{B}}
\newcommand{\W}[0]{W}
\newcommand{\F}[0]{\mathcal{F}}
\newcommand{\I}[0]{\mathsf{I}}

\title[CRAFT]{An Iterative Reconstruction Algorithm for Faraday Tomography}

\author[S. Cooray et al.]{Suchetha Cooray,$^{1}$\thanks{E-mail: \href{mailto:cooray@nagoya-u.jp}{cooray@nagoya-u.jp}}
Tsutomu T. Takeuchi,$^{1,2}$
Takuya Akahori,$^{3}$
Yoshimitsu Miyashita,$^{4}$
\newauthor  Shinsuke Ideguchi,$^{5}$
Keitaro Takahashi,$^{4, 6}$
 and Kiyotomo Ichiki$^{1, 7}$
\\
\\
$^{1}$ Division of Particle and Astrophysical Science, Nagoya University, Furo-cho, Chikusa-ku, Nagoya 464–8602, Japan\\
$^{2}$ The Research Center for Statistical Machine Learning, The Institute of Statistical Mathematics, 
10-3 Midori-cho, Tachikawa, Tokyo 190-8562, Japan\\
$^{3}$ Mizusawa VLBI Observatory, National Astronomical Observatory of Japan, 2-21-1 Osawa, Mitaka, Tokyo 181-8588, Japan\\
$^{4}$ Kumamoto University, 2-39-1, Kurokami, Kumamoto 860-8555, Japan\\
$^{5}$ Department of Astrophysics/IMAPP, Radboud University Nijmegen, PO Box 9010, NL-6500 GL Nijmegen, the Netherlands\\
$^{6}$ International Research Organization for Advanced Science and Technology, Kumamoto University, Japan\\
$^{7}$ Kobayashi-Maskawa Institute for the Origin of Particles and the Universe, Nagoya
University, Chikusa-ku, Nagoya, 464-8602, Japan
}

\date{Accepted XXX. Received YYY; in original form ZZZ}

\pubyear{2020}

\begin{document}
\label{firstpage}
\pagerange{\pageref{firstpage}--\pageref{lastpage}}
\maketitle

\begin{abstract}
    
    Faraday tomography offers crucial information on the magnetized astronomical objects, such as quasars, galaxies, or galaxy clusters, by observing its magnetoionic media. The observed linear polarization spectrum is inverse Fourier transformed to obtain the Faraday dispersion function (FDF), providing us a tomographic distribution of the magnetoionic media along the line of sight. However, this transform gives a poor reconstruction of the FDF because of the instrument's limited wavelength coverage. The current Faraday tomography techniques' inability to reliably solve the above inverse problem has noticeably plagued cosmic magnetism studies. We propose a new algorithm inspired by the well-studied area of signal restoration, called the {\sl Constraining and Restoring iterative Algorithm for Faraday Tomography} (CRAFT). This iterative model-independent algorithm is computationally inexpensive and only requires weak physically-motivated assumptions to produce high fidelity FDF reconstructions. We demonstrate an application for a realistic synthetic model FDF of the Milky Way, where CRAFT shows greater potential over other popular model-independent techniques. The dependence of observational frequency coverage on the various techniques' reconstruction performance is also demonstrated for a simpler FDF. CRAFT exhibits improvements even over model-dependent techniques (i.e., QU-fitting) by capturing complex multi-scale features of the FDF amplitude and polarization angle variations within a source. The proposed approach will be of utmost importance for future cosmic magnetism studies, especially with broadband polarization data from the Square Kilometre Array and its precursors. We make the CRAFT code publicly available\footnotemark.
    
\end{abstract}


\footnotetext{{\url{https://github.com/suchethac/craft}}}
\section{Introduction} \label{sec:introduction}

    Cosmic magnetism influences a wide range of astrophysical phenomena from interstellar gas to galaxy clusters. Thus, understanding cosmic magnetism has become a key science goal for the present and future radio telescopes \citep[e.g.,][]{Gaensler_2004, Beck_2009, Akahori_2016, Akahori_2018a}. The most promising approach to measure the magnetic field strength in the Universe is to observe the polarized synchrotron radiation. When linearly polarized emission passes through a thermal magnetized plasma (magnetoionic media), the polarization angle undergoes a frequency-dependent rotation, which can be used to trace the cosmic magnetic fields along the line-of-sight (LOS) \citep{Kronberg_1982, Kolatt_1998, Stasyszyn_2010, Akahori_2014}. 
    
    With the advancement of broadband radio polarimetry, Faraday rotation measure (RM) synthesis \citep{Burn_1966, Brentjens_2005, Heald2009} has become an indispensable tool for analyzing multichannel polarization data. RM synthesis converts the linear polarization spectrum to the Faraday dispersion function (FDF), a distribution of polarized intensity as a function of Faraday depth. The FDF constitute the Faraday rotation measure (RM), thermal gas density-weighted ﬁeld strength along the LOS, and provide tomographic information of polarized emitters and Faraday rotating bodies (plasma) across Faraday depth (see Section \ref{sec:faraday_tomography} for details). Obtaining tomographic information of the magnetic fields by RM synthesis is often known as Faraday tomography.
    
    However, Faraday tomography is complicated by limited polarization observation in wavelength space (more precisely wavelength squared), and reconstructing the intrinsic FDF from incomplete information is a well-known challenge \citep[e.g.,][]{Beck_2012, Akahori_2014}. Many techniques have been proposed for FDF reconstruction. Nevertheless, they suffer from a subpar performance in the FDF reconstruction accuracy because of limited polarization observations in frequency \citep{Andrecut_2012, Kumazaki_2014, Sun_2015}. 
    
    A possible Faraday tomography technique is fitting parameterized models to the observed polarization spectrum when the shape of the FDF can be known or assumed. The fitting, often known as QU-fitting, can be done using the method of least square \citep{Farnsworth_2011, O'Sullivan_2012, Ideguchi_2014a, Ozawa_2015, Kaczmarek_2017}, or by Markov Chain Monte Carlo (MCMC) approaches \citep{Sakemi_2018, Schnitzeler_2018, Miyashita_2019}. QU-fitting can recognize overlapping components and is shown to perform better in some situations, such as in low signal-to-noise conditions \citep{Sun_2015}. Under the appropriate circumstance, the QU-fitting problem is relatively simple, requiring only to estimate the parameters of one or a mixture of a few analytic functions (e.g., Gaussian, top-hat, and delta functions).
    
    On the other hand, a possible model-independent Faraday tomography technique is by assuming the sparsity of the reconstructed FDF. Faraday tomography's mathematical formalism is similar to that of radio interferometric imaging (e.g., \citet{Thompson_2017}), where sparsity is often employed for image reconstruction. There are two radio interferometric imaging techniques applied to Faraday tomography. First is the classical technique of CLEAN \citep{Hogbom_1974}, a matching pursuit algorithm that is implemented to Faraday tomography as RM CLEAN \citep[e.g.,][]{Heald_2009, Anderson_2016, Michilli_2018}. However, the CLEAN algorithm is known to be poor at reconstructing extended sources in the image space, though some workarounds have been studied \citep[e.g.,][]{Cornwell_2008}. The second technique solves the observational equation (Eq. (\ref{eq:mask})) with ideas based on compressed-sensing theory \citep{Donoho_2006, Candes_2006}. They rely on regularization functions to select the sparse solution from the infinitely many solutions. \cite{Li_2011, Andrecut_2012}, and recently \citet{Akiyama_2018} has shown the use of sparsity regularized reconstruction techniques in Faraday tomography. Solving these regularized optimization problems can be extremely computationally expensive even for simple Faraday structures as the complexity of the problem (number of parameters to estimate) can quickly increase with higher resolution or wider domain of FDF.
    
    Faraday tomography is challenging, as seen above because it can be classified as an \textit{inverse problem}, where we try to reconstruct the complete signal from a distorted one with some known/assumed information. The main challenge of inverse problems is that they are often also \textit{ill-posed}, meaning that infinitely many solutions exist for an observation. Solving inverse problems for signals is a well-studied area called multidimensional signal restoration \citep{Dudgeon_1984}, which suggests that the solution to an inverse problem can be found iteratively by successive approximations while imposing assumed constraints about the signal. This iterative signal restoration technique has been successfully applied to deconvolution \citep{Schafer1981, Mersereau_1978, Richards_1979}, super-resolution \citep{Gerchberg1974}, signal extrapolation \citep{Landau_1961,Papoulis1975}, and denoising \citep{Frieden_1975,Fienup_1978}. 
    
    Due to FDF reconstruction techniques' current limitations, it is important to explore different Faraday tomography approaches. Using the finite domain in Fourier space as an assumption, \citet{Cooray_2020} showed the use of the iterative signal restoration technique for partial astronomical signals. By the assumption that the intrinsic FDF is limited in Faraday depth, the formalism discussed in \citet{Cooray_2020} can be extended to Faraday tomography.
    
    This paper introduces a novel reconstruction technique for Faraday tomography called CRAFT (Constraining and Restoring iterative Algorithm for Faraday Tomography) and is structured as follows. In Section \ref{sec:faraday_tomography}, we explain the basics of Faraday tomography and its necessary mathematical formulation. Next, in Section \ref{sec:technique}, we introduce our CRAFT technique. We show an example reconstruction of a realistic galactic model simulated FDF in Section \ref{sec:application}, followed by the dependence of observational frequency coverage on CRAFT and the currently available techniques in Section \ref{sec:spectral_dependence}. We present some comments in Section \ref{sec:discussion}, including concerns for observations. Lastly, in Section \ref{sec:conclusion}, we summarize this work with future prospects.

\section{Faraday Tomography} \label{sec:faraday_tomography}

    Polarized emission can be described using the four Stokes parameters, $I$, $Q$, $U$, and $V$. Stokes $I$ represents the total intensity, $Q$ and $U$ represent the two components of linear polarization, and $V$ represents the circular polarization. For Faraday tomography, we are interested in the linear polarization components and are combined to give the complex linear polarization spectrum $P = Q + iU$. The complete polarization spectrum can be defined as;
    \begin{equation}
        P\left(\lambda^{2}\right)=\int_{0}^{\infty} \varepsilon(r) e^{2 i \chi\left(r, \lambda^{2}\right)} d r=\int_{-\infty}^{\infty} F(\phi) e^{2 i \phi \lambda^{2}} d \phi ,
    \end{equation}
    where $\lambda$ is the wavelength of the emission, $\varepsilon$ is the synchrotron polarization emissivity along the LOS, $\phi$ is Faraday depth, which is proportional to the integration of thermal electron density and magnetic ﬁelds along the LOS, and $F(\phi)$ is the Faraday dispersion function \citep[FDF;][]{Burn_1966, Brentjens_2005}. The above equation satisfies the form of Fourier transform, hence the FDF can be obtained by the inverse Fourier transform of the complex linear polarization spectrum $P(\lambda^2)$.
    
    However, in real observations, the wavelength coverage is finite, and negative $\lambda$ is nonphysical. Thus, the observed complex linear polarization spectrum $\tilde{P}(\lambda^2)$ is written as,
    \begin{equation} \label{eq:mask}
    \tilde{P}\left(\lambda^{2}\right)= \W \left(\lambda^{2}\right) P\left(\lambda^{2}\right), 
    \end{equation}
    where $\W(\lambda^2)$ is the sampling function and is nonzero where there is linear polarization measurement. The Fourier transform of this sampling function $\W(\lambda^2)$ is often known as the rotation measure spread function (RMSF). The limited information of the observed spectrum implies that Faraday tomography is an ill-posed problem. Moreover, the knowledge of the intrinsic FDF is required to obtain the correct solution from the infinite possible ones that satisfy the observed spectrum. Successful reconstruction is possible with a prior knowledge of the expected FDF, given the sources are detected despite Faraday depolarization \citep[e.g.,][]{Farnsworth_2011}.

\section{Reconstruction Technique} \label{sec:technique}

    In this work, we propose a reconstruction technique for Faraday tomography based on iterative signal restoration. Our final goal is to reconstruct the full linear polarization spectrum $P(\lambda^2)$ from the observed spectrum $\tilde{P}(\lambda^2)$. In other words, we explore the inverse of the sampling function, $\W^{-1}$, of the following inverse equation,
    \begin{equation} \label{eq:inverse}
    {P}\left(\lambda^{2}\right)= \W^{-1}\left(\lambda^{2}\right) \tilde{P}\left(\lambda^{2}\right).
    \end{equation}
    To solve this ill-posed inverse problem, one needs to regularize the problem by assuming constraints on the reconstruction. 
    
    We propose an iterative algorithm for the above, in which successive approximations are made to get to a better estimate every time. At each iteration, assumptions about the underlying FDF are imposed on the estimated FDF. For example, we can consider that some parts of the FDF from partial observations ("dirty" FDF) to be purely a result of the RMSF. In such a case, limiting the domain of the FDF in $\phi$ can remove some fringe effects of the RMSF. The corrected FDF is then Fourier transformed to obtain an estimate of the linear polarization spectrum. This estimated linear polarization spectrum is nonzero even in the spectrum's unobserved parts due to the a priori information imposed on the FDF transformed. The observed region of the linear polarization spectrum is then restored by combining it with the previously estimated spectrum. The new linear polarization spectrum is inverse Fourier transformed to obtain the next estimate of the FDF. The latest estimate is a better estimate of the FDF than the previous, which is repeated until convergence. Restoring the observed spectrum ensures the integrated intensity is conserved, and repeating the above-explained procedure will produce the desired FDF.
    
    Let $\B$ be the operator that contains the a priori knowledge of $P$ such that when $\B$ operates on a spectrum, constraints are imposed on the estimation at each iteration. Mathematically we have the equation,
    \begin{equation}
        P^{\prime} \left(\lambda^{2}\right) = \B P \left(\lambda^{2}\right),
    \end{equation}
    where $P^{\prime}$ is the spectrum that obeys the constraints encapsulated in $\B$ about the intrinsic signal. The first constraint of $\B$ can incorporated as $\B = \F \mathsf{\beta} \F^{-1}$ where $\F$ is an operator of Fourier transform, and $\mathsf{\beta}$ is a window function in Faraday depth space. Initially, we can confine the non-zero parts of the FDF to a finite Faraday-depth range that is physically-motivated for the astronomical target. The FDF confinement in Faraday depth is also natural as Faraday depth accumulates like a random walk process and do not reach large values \citep{Ideguchi_2014b}. We could also impose sparsity in Faraday-depth by incorporating a nonlinear threshold operator $S_{\mu}$ \citep{Daubechies_2004,Kayvanrad_2009} within $\B$ as,
    \begin{equation} \label{eq:thresholding_operator}
        S_{\mu}(|F(\phi)|)=\left\{
        \begin{array}{ll}
            |F(\phi)|-\mu & \text { if } |F(\phi)| \ge \mu \\
            0 & \text { if }|F(\phi)| < \mu \\
        \end{array}\right. ,
    \end{equation}
    where $\mu$ acts like a soft cutoff for the FDF amplitude. Additionaly, we explore the smoothing of the polarization angle as a possible constraint for Faraday tomography (see Section \ref{sec:application}).
    
    The above explained reconstruction technique for the $n^{\textrm{th}}$ iteration can be simply written as;
    \begin{equation} \label{eq:algorithm}
        P_{n} (\lambda^{2}) = \tilde{P} (\lambda^{2}) + \left[\I - \W (\lambda^{2}) \right] \B P_{n-1}  (\lambda^{2}),
    \end{equation}
    where $\I$ is the identity matrix, $P_{n}$ is the $n^{\textrm{th}}$ estimate of $P$, and $P_0 = \tilde{P}$. In the above equation, the second term represents the $n^{\textrm{th}}$ guess of the missing polarization spectrum, which is combined with the observed spectrum to produce $P_{n}$.
    
    This algorithm is a version of projected gradient descent \citep{Combettes_2009} and therefore as $n$ tends to infinity, the solution/estimate will converge towards the original spectrum $P$. That is,
    \begin{equation} \label{eq:convergence}
        P_{n} \rightarrow P \textrm{ as } n \rightarrow \infty .
    \end{equation}
    The iteration is terminated when the residual between the successive estimates ($|P_{n}-P_{n-1}|$) are infinitesimal. Practically, the stopping criterion we use is $||P_{n}-P_{n-1}||/||P_{n-1}||<\epsilon$, where $\epsilon$ is some small positive number. That said, it should be noted that the degree of possible reconstruction is determined by observation and the constraints we impose. We find that the spectrum is reconstructed fairly well for $|\lambda^2| \le \lambda^2_{\mathrm{max}}$, where $\lambda^2_{\mathrm{max}}$ is the maximum observed $\lambda^2$ (see Section \ref{sec:application}, for details). The reconstructed FDF is then just the inverse Fourier transform of the estimated linear polarized spectrum.

\section{Reconstructing a Realistic Synthetic Spectrum} \label{sec:application}
    
    We show a demonstration of the proposed method for a synthetic FDF described in \citet{Ideguchi_2014b} of a sophisticated model simulation of the Milky Way \citep{Akahori_2013}. The synthetic model FDF contains complicated structures that include both Faraday-thin ($\lambda^2 \Delta \phi \ll 1$) and thick ($\lambda^2 \Delta \phi \gg 1$) components, where $\Delta \phi$ is the extent of the source in $\phi$. Such a complicated FDF will be too difficult to approximate by analytic functions in model fitting techniques (e.g., QU-fitting) and will not be considered for comparison.
    
    \begin{figure}
        \centering
        \includegraphics[width=0.95\linewidth]{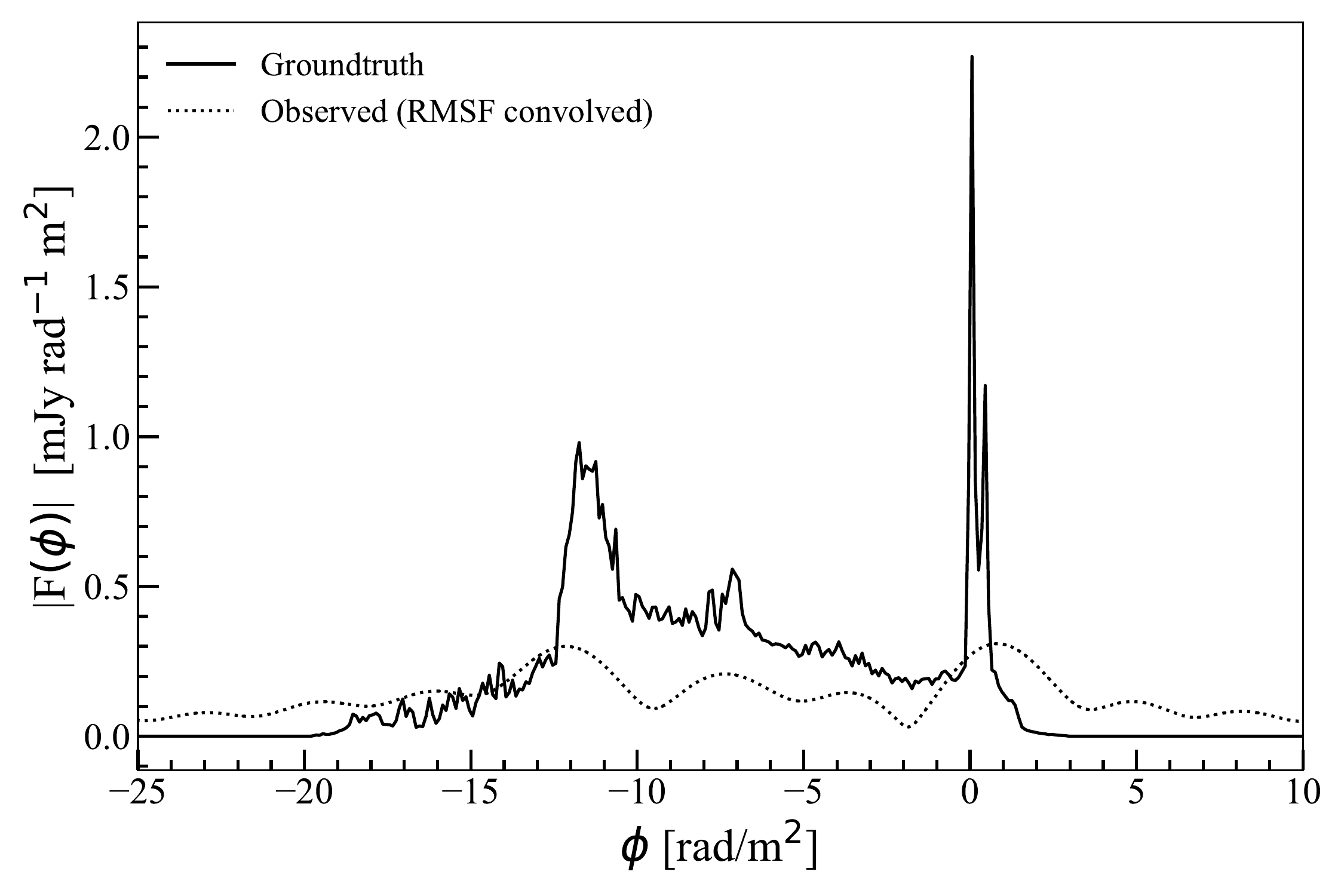}
        \caption{Synthetic simulation of a FDF \citep{Ideguchi_2014b} of a complex model of the Milky Way as described in \citet{Akahori_2013} (solid line) and the RMSF convolved FDF due to limited $\lambda^2$ coverage that corresponds to a frequency coverage of 300 [MHz] to 3000 [MHz] (dotted line).}
        \label{fig:fdf_model}
    \end{figure}
    
    The model FDF is the volume integrated FDF simulation for a Milky Way-like galaxy with a sophisticated galactic model, incorporating Magneto-hydrodynamic turbulence \citep[see][for details.]{Ideguchi_2014b}. The model FDF has a Faraday depth range of -1000 to 1000 [rad m$^{-2}$] and a grid size of 0.1 [rad m$^{-2}$], and the polarization spectrum is obtained by numerically Fourier transforming the model FDF. The polarization observations were of {630} channels of frequency between 300 [MHz] and 3000 [MHz]. The frequency range is the optimum range for exploring intergalactic magnetism \citep{Akahori_2018b}, and the same as that is used in \citet{Akiyama_2018}, allowing us to compare our results with theirs directly. For simplicity, each channel is assumed to be infinitesimally narrow and spaced equally in $\lambda^2$ space. We add a random Gaussian noise of zero mean and a standard deviation of 0.1 [mJy] to the $Q$ and $U$ measurements separately in each channel. The RMSF for the above setup has a full width at half maximum (FWHM) of $2\sqrt{3}/(\lambda^2_{\textrm{obs, max}} - \lambda^2_{\textrm{obs, min}}) = 3.50$ [rad m$^{-2}$], where $\lambda^2_{\textrm{obs, min}}$ and $\lambda^2_{\textrm{obs, max}}$ are the minimum and the maximum $\lambda^2$ values in the observation coverage. The model used for this experiment and the RMSF-convolved (observationally-available) FDF due to the limited $\lambda^2$ coverage is shown in Figure \ref{fig:fdf_model}.
    
    \begin{figure*}
        \centering
        \includegraphics[width=0.95\textwidth]{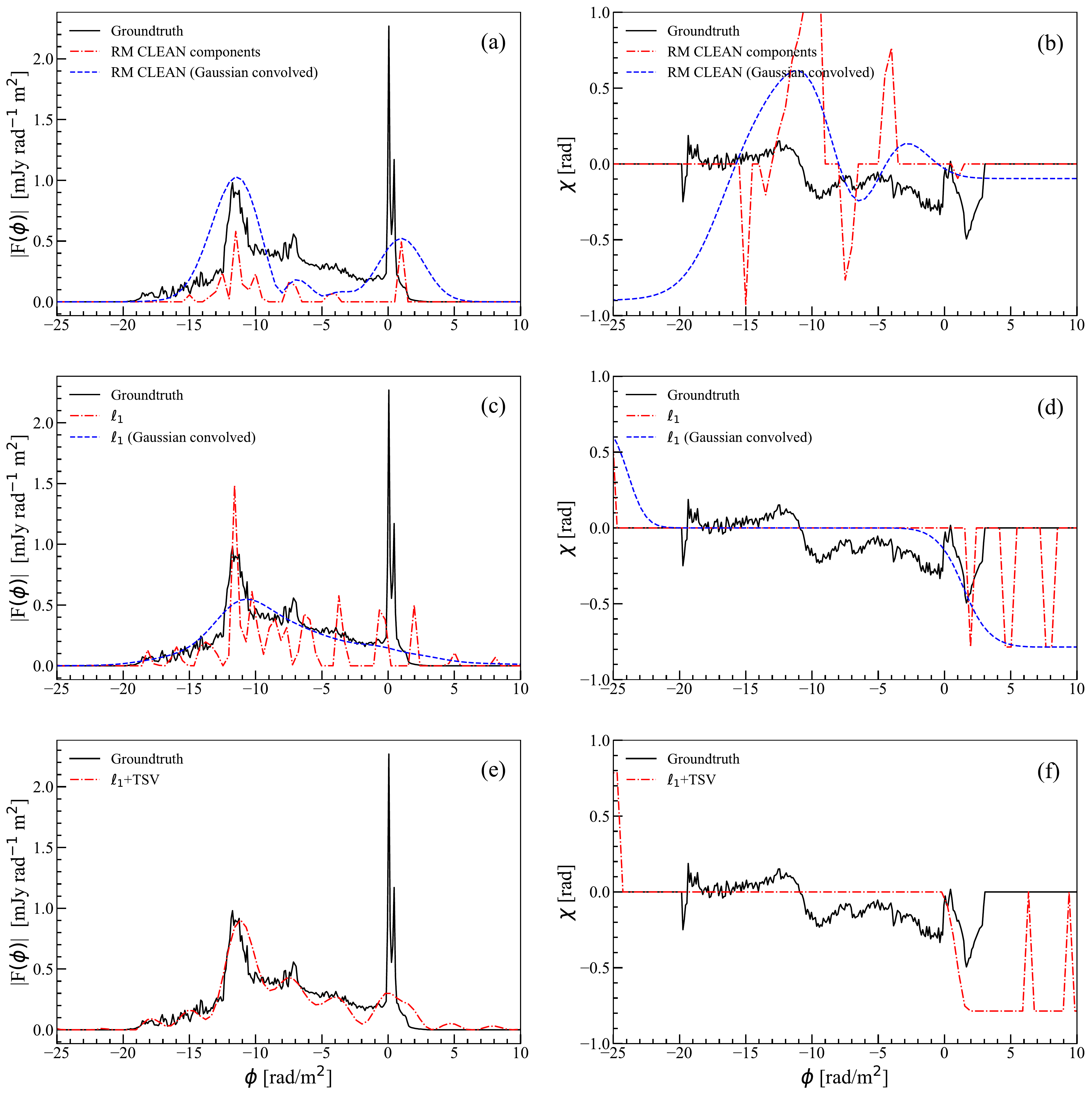}
        \caption{The figure shows the reconstruction capability by the popular existing Faraday tomography techniques for the setup explained in Section \ref{sec:application}. The left panels (a), (c), (e) shows the amplitudes of the FDF reconstructions by RM CLEAN, $\ell_1$ regularized reconstruction, and $\ell_1$+TSV regularized reconstruction, respectively. On the other hand, the right panels (b), (d), (f) are the corresponding polarization angles for the left panels. The solid black line shows the original model FDF and the red dash-dotted line show the reconstructed FDF by those techniques. A Gaussian kernel smoothes the obtained FDFs by RM CLEAN and $\ell_1$ regularized reconstruction with the FWHM equivalent to that of the RMSF. The smoothed FDFs, and corresponding polarization angle is shown by the blue dashed line.}
        \label{fig:recons_comparison}
    \end{figure*}
    
    Figure \ref{fig:recons_comparison} shows the amplitudes and the polarization angles of the reconstructed synthetic FDFs by existing model-independent methods. From top to bottom of the figure, we show the best reconstructions using RM CLEAN \citep{Heald2009} implemented in \citet{Miyashita_2016} with grid size of 0.5 [rad m$^{-2}$], gain of 0.1, and threshold 0.1 [mJy rad$^{-1}$ m$^2$], $\ell_1$ sparse reconstruction \citep{Li_2011} with $\Lambda_{\ell} = 10^{-1}$, and $\ell_1$ + total square variance (TSV) sparse reconstruction \citep{Akiyama_2018} with $(\Lambda_{\ell}, \Lambda_{t}) = (10, 10^3)$. The two latter methods used a grid size of 0.43 [rad m$^{-2}$]. Reconstructions with RM CLEAN and $\ell_1$ regularization deviate widely from the groundtruth and show that sparsity in $\phi$ is not suited for extended FDFs with Faraday-thick components. $\ell_1+$TSV regularized reconstruction allows for smoother extended reconstruction by nature, allowing for better reconstruction of the synthetic FDF. However, we observe that the $\ell_1+$TSV regularized technique fails to simultaneously recover multi-scale features. The cross-validation selected parameters of $\Lambda_{\ell}$ and $\Lambda_{t}$ for this FDF favors the extended components, compromising the Faraday-thin peak at $\phi \approx 0-1$ [rad m$^{-2}$].
    
    The constraint operator on the spectrum for our reconstruction technique is $\B = \F \mathsf{\beta} \F^{-1}$ where $\mathsf{\beta}$ is a window function in Faraday depth space. The window is determined by physical constraints on possible Faraday depth $\phi$. In this work, we initially set a very loose constraint of $|\phi| \le 500$ [rad m$^{-2}$]. This constraint is suitable for the majority of the observed rotation measures in all-sky rotation measure surveys \citep[e.g.,][]{Taylor_2009}. The window in $\phi$ is updated using the estimate of the FDF at each iteration. Additionally, we use the non-linear thresholding operator $S_{\mu}$ defined in Eq. (\ref{eq:thresholding_operator}) with $\mu=0.01$. The allowed $\phi$ values are constrained to $|F_n(\phi)| \ge \mu$ and otherwise set to zero. The threshold parameter $\mu$ can be selected through a grid search, as using inappropriate parameters will disturb the convergence to a solution. The iteration stopping criterion was $\epsilon = 0.001$. 
    
    The unattainable nature of the observed polarization spectrum for negative $\lambda^2$ makes the determination of phase information particularly difficult in Faraday tomography. In the above-explained technique, all the Faraday components are reproduced in the resultant FDF. However, we observed that the reconstruction of the polarization angle and the negative $\lambda^2$ side of the polarization spectrum were poor. These reconstruction results are shown in Appendix \ref{sec:appendix_a}.
    
    A possible improvement strategy is to assume some physical properties of the observed magnetoionic media. \citet{Frick_2010} has demonstrated a technique by reasoning the astronomical object's symmetry along the LOS and using wavelet transforms to decompose the multi-scale components in Faraday depth space. However, for complex, mixed, and asymmetric FDFs, such techniques are difficult to apply.
    
    As a possible solution, we propose to smooth the polarization angle in $\phi$ at each iteration. When we attempt to reconstruct the whole $\lambda^2$ domain (including the negative $\lambda^2$), the RMSF rotates rapidly within its main lobe because we consider $\lambda^2_0=0$ in the Eq. (25) and (26) of \citet{Brentjens_2005}. Thus, we argue that neglecting scales in $\phi$ that are smaller than RMSF's main lobe width should negate this effect. The smoothing can be done by a Gaussian kernel of FWHM = $2\pi/(\lambda^2_{\textrm{obs, max}} - \lambda^2_{\textrm{obs, min}}) = 6.36 $ [rad m$^{-2}$], which is the RMSF main lobe width. Effectively, we compromise the resolution of $\chi$ for accurate reconstructions at scales larger than RMSF's main lobe width. Precisely reconstructing the phase for scales smaller than the RMSF main lobe width requires the impossible observations in negative $\lambda^2$ or physically motivated assumptions based on the astronomical target. However, by Gaussian smoothing the polarization angle and neglecting the smaller scales, the algorithm converges to an accurate smoothed reconstruction of $\chi$. We note that a Gaussian kernel is used because it does not introduce artifacts when convolved repeatedly.

    \begin{figure*}
        \centering
        \includegraphics[width=0.7\linewidth]{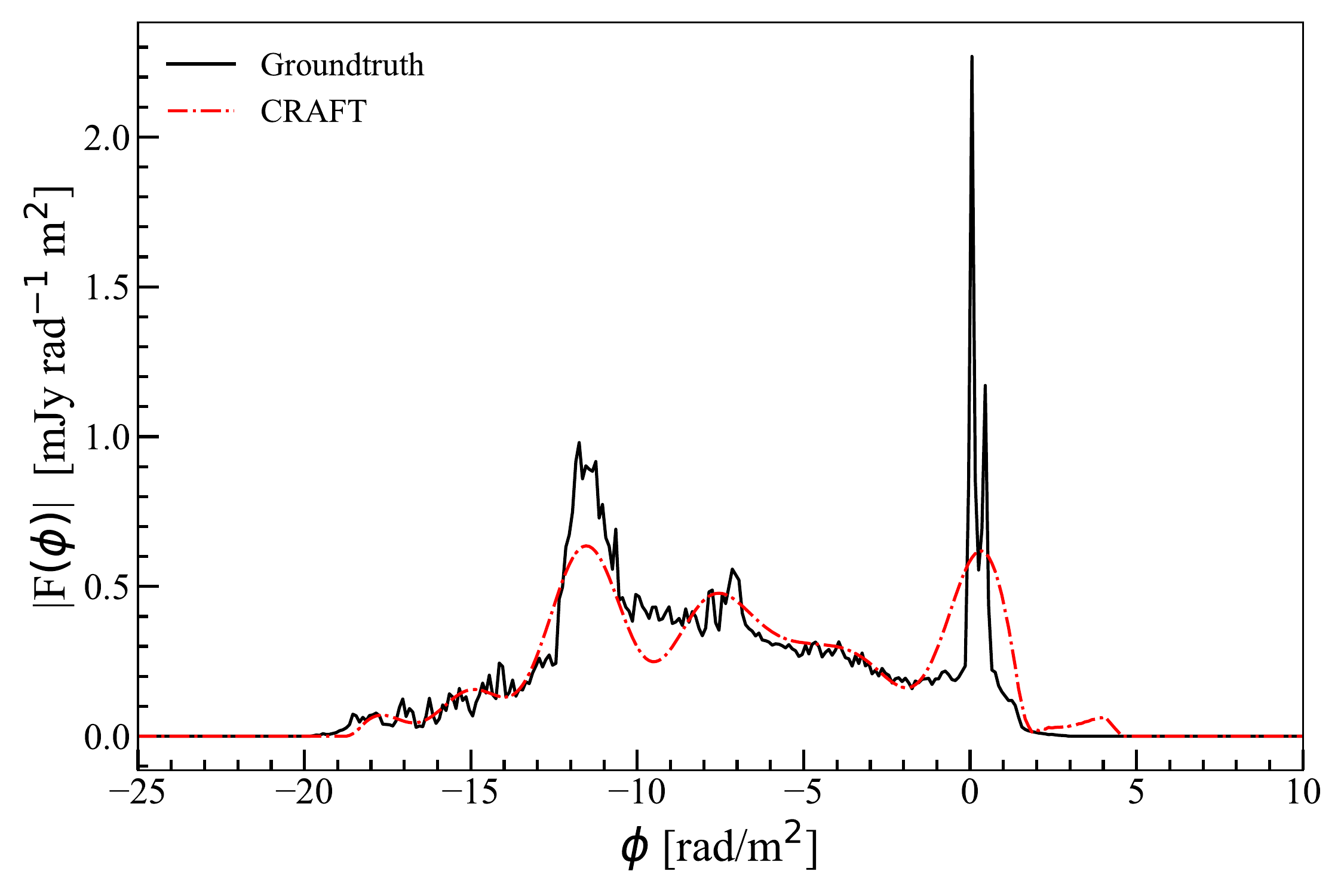}
        \caption{Amplitude  of the realistic galaxy FDF reconstruction by CRAFT with observations between 300 [MHz] to 3000 [MHz]. Black solid line show the original model FDF and the red dash dotted line show the reconstructed FDF.}
        \label{fig:reconstruction_amplitude}
    \end{figure*}
    \begin{figure}
        \centering
        \includegraphics[width=0.95\linewidth]{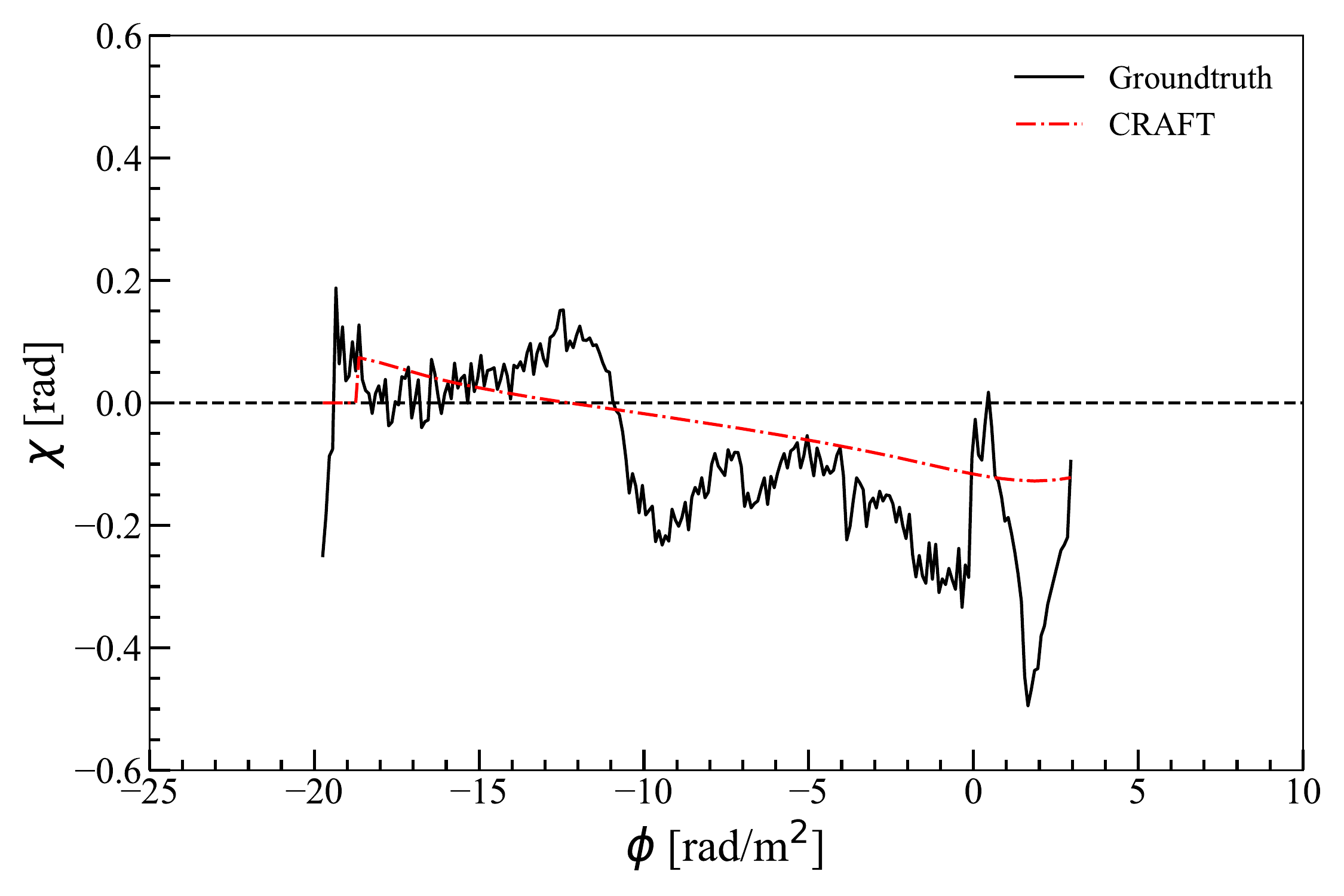}
        \caption{Polarization angle of the realistic galaxy FDF reconstruction by the proposed technique with observations between 300 [MHz] to 3000 [MHz]. Black solid line show the synthetic simulation and the red dash dotted line show the reconstructed.}
        \label{fig:reconstruction_phase}
    \end{figure}
    
    The FDF reconstruction with this procedure is shown in Figure \ref{fig:reconstruction_amplitude}. The stopping criterion was reached in 532 iterations with the final $\phi$ window approximately between -19 [rad m$^{-2}$] and 5 [rad m$^{-2}$]. The results are promising, producing an FDF with both Faraday-thin and Faraday-thick components properly reconstructed. The reconstructed polarization angle is shown in Figure \ref{fig:reconstruction_phase}. The reconstructed polarization angle is in good agreement, despite the compromised resolution for scales smaller than 6.36 [rad m$^{-2}$]. We note that as a consequence of smoothing, the region of the FDF between -13 [rad m$^{-2}$] and -8 [rad m$^{-2}$] is poorly reconstructed due to the quick (scales smaller than 6.36 [rad m$^{-2}$]) change in the polarization angle at around -11 [rad m$^{-2}$]. Regardless, the large scale accuracy in the reconstructed polarization angle is beyond any of the techniques tested in this paper.

    \begin{figure}
        \centering
        \includegraphics[width=0.95\linewidth]{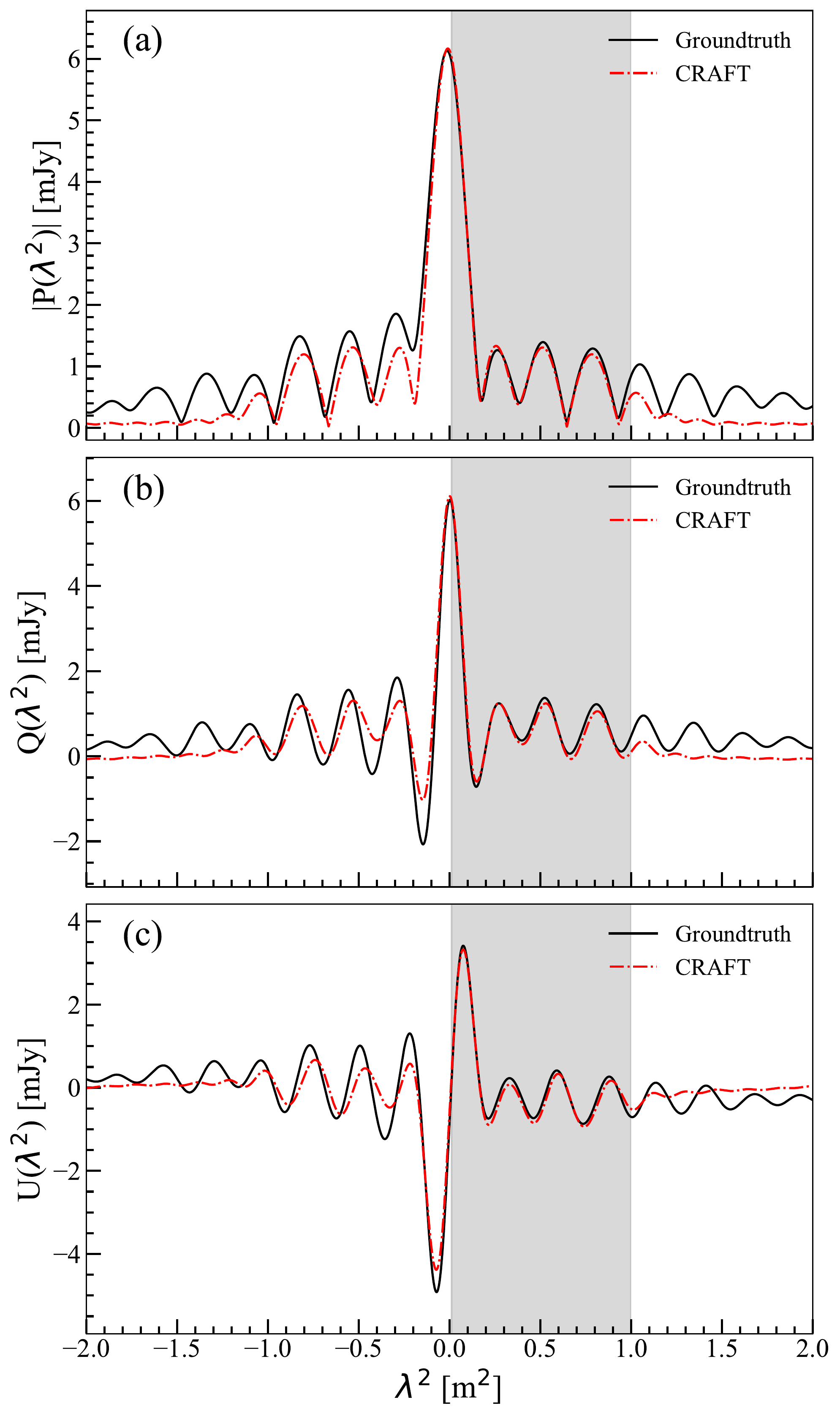}
        \caption{CRAFT reconstructed linear polarization spectrum of the realistic galaxy FDF. Panel (a) shown the amplitude of the complex linear polarization spectrum, (b) is the  Stokes Q, and (c) is the Stokes U. The black dotted line is the of the noiseless synthetic model FDF and the red dash dotted line show its reconstruction by the proposed technique. Observed $P(\lambda^2)$ are approximately between 0.01 [m$^2$] and 1.00 [m$^2$], shown by the shaded region.}
        \label{fig:reconstruction_spectrum}
    \end{figure}
    
    The reconstructed polarization spectrum is shown in Figure \ref{fig:reconstruction_spectrum}. The spectrum is well reproduced even on the negative $\lambda^2$ side, at least to $|\lambda^2| < \lambda^2_{\textrm{obs, max}}$, after which the amplitude tapers off with increasing $|\lambda^2|$. Tapering off suggests the lack of information at smaller scales in $\phi$ for the reconstruction algorithm.
    
    For quantitative analysis of the reconstruction performance between the techniques, we suggest the use of normalized root mean square error \citep[NRMSE: e.g.,][]{Fienup_1997}. NRMSE for FDF reconstructions can be defined by,
    \begin{equation}
        \textrm{NRMSE}\left(\hat{F}, {F}\right) = \sqrt{ \frac {\sum_{i}\left| \hat{F}_{i}-{F}_{i} \right|^{2} } {\sum_{i} \left| {F}_{i} \right|^{2}}} ,
    \end{equation}
    where $\hat{F}$ and $F$ are reconstructed and the model complex FDFs, respectively. The reconstruction accuracy indicated by the NRMSE of various methods discussed above is shown in Table \ref{table:nrmse}. As shown, the proposed technique outperforms the previous techniques with an NRMSE of 0.46. If we assume that the reconstruction of the linear polarization spectrum is possible between $-\lambda^2_{\textrm{obs, max}}$ and $+\lambda^2_{\textrm{obs, max}}$, the $\Delta \lambda^2$ term of Eq. (61) in \citet{Brentjens_2005} becomes $2\lambda^2_{\textrm{obs, max}}$. Therefore, the smallest possible $\phi$ scale for reconstruction can be written as,  
    \begin{equation}\label{eq:recon_scale}
        \delta \phi_{\textrm{reconstructed}} = \frac{\sqrt{3}}{\lambda^2_{\textrm{obs, max}}}.
    \end{equation}
    As a comparison, we smooth the synthetic model FDF with a Gaussian kernel of FWHM = 1.73 [rad m$^{-2}$] (using Eq. \ref{eq:recon_scale}), which results in an NRMSE of 0.46. The reconstruction of CRAFT is comparable to the smoothed model FDF, indicating great potential of CRAFT. In Appendix \ref{sec:appendix_b}, we show the synthetic model FDF smoothed at different scales to be compared with the reconstruction.
    
    \begin{table}
        \centering
        \begin{tabulary}{\linewidth}{l C} 
            \hline
            Reconstruction Method & NRMSE\\
            \hline \hline
            RM CLEAN \footnotemark[1] & {1.31} \\
            $\ell_1$ regularization \footnotemark[1] & {0.58} \\
            $\ell_1$+TSV regularization & {0.53} \\
            CRAFT (This work) & \textbf{{0.46}} \\ 
            \hline
        \end{tabulary}
        \caption{A comparison of the reconstruction performances for the synthetic FDF described in \citep{Ideguchi_2014b} for a observed frequency coverage of 300 [MHz] to 3000 [MHz].}
        \label{table:nrmse}
    \end{table}
    \footnotetext[1]{The result is after post-processing by Gaussian convolution with FWHM $\equiv$ RMSF FWHM = 3.50 [rad m$^{-2}$], which is often performed in radio interferometry imaging \citep[e.g.,][]{Thompson_2017}.}
\section{Dependence on observational frequency coverage} \label{sec:spectral_dependence}

    By focusing on a typical extragalactic observation, we investigate the dependence of observational frequency coverage on the different Faraday tomography methods' reconstruction performance. A simple analytic FDF model is considered for possible observations by currently available Australian Square Kilometre Array Pathfinder \cite[ASKAP;][]{ASKAP} and the upcoming Square Kilometre Array (SKA) Phase 1 mid bands 1 and 2 instrument (hereafter simply, SKA1 mid).
    
    The FDF model is a combination of a diffuse source and a compact source separated in $\phi$, as also done in \citet{Akahori_2014}. The diffuse component is modeled as,
    \begin{equation}
        \begin{aligned} F_{\mathrm{d}}(\phi) = F_{\mathrm{d} 0} &\left\{ \frac{1}{4}\left[\tanh \left(\pi \frac{\phi-\phi_{\mathrm{dw}}-\phi_0} {\phi_{\mathrm{dw}}}\right) \right] \right.\\ &\left.\times \left[1 + \tanh \left(-\pi \frac{\phi-\phi_{\mathrm{d}}-\phi_{\mathrm{dw}}-\phi_0} {\phi_{\mathrm{dw}}} \right)\right] \right\}, \end{aligned}
    \end{equation}
    where $F_{\mathrm{d}0}$ is the normalization constant of the amplitude, $\phi_0$ is the position of the left edge of the source, $\phi_{\mathrm{dw}}$ is the width of the tails, and $\phi_{\mathrm{d}}$ is the width of the flat section of the source. The diffuse source is modeled based on the hyperbolic tangent because of the platykurtic proﬁles produced by the simulated Galaxy models \citep{Sun_2008,Waelkens_2009, Akahori_2013,Ideguchi_2014b}. For this diffuse component, the parameters are set as $F_{\mathrm{d}0} = 0.5$ [mJy], $\phi_0=-15$ [rad m$^{-2}$], $\phi_{\mathrm{dw}} = 2$ [rad m$^{-2}$], and $\phi_{\mathrm{d}}=6$ [rad m$^{-2}$].
    
    The compact source, such as from a quasar or a radio galaxy, is modeled by a Gaussian \citep{Burn_1966,Frick_2010} as;
    \begin{equation}
        F_{\mathrm{c}}(\phi)=F_{\mathrm{c} 0} \exp \left\{-\frac{\left(\phi-\phi_{\mathrm{c}}\right)^{2}}{2 \phi_{\mathrm{cw}}^{2}}\right\} ,
    \end{equation}
    where $F_{\mathrm{c} 0}$ is the normalization constant, $\phi_{\mathrm{c}}$ is the location of the peak in $\phi$ space, and $\phi_{\mathrm{cw}}$ characterizes the width of the Gaussian. For the compact source component of our model, $F_{\mathrm{c} 0}=1$ [mJy], $\phi_{\mathrm{c}}=10$ [rad m$^{-2}$], $\phi_{\mathrm{cw}}=0.2$ [rad m$^{-2}$]
    
    We consider the intrinsic polarization angle $\chi$ to be independent of $\phi$ and equal to zero. Assuming a constant $\chi$ is in stark contrast to the highly varying polarization angle of the previous model, and enables us to solely investigate the frequency coverage dependence on reconstruction performance (especially on the polarization angle). A situation with $\chi=0$ also suggests that the FDF is purely real. Thus, ideally, for this case, sparsity regularized reconstructions would suppress all the imaginary components of the FDF to produce good reconstructions. In this setup, we posit that deviations in the reconstructed polarization angle can be considered artifacts and affect the astrophysical interpretation of a reconstructed FDF.
    
\begin{figure}
        \centering
        \includegraphics[width=0.95\linewidth]{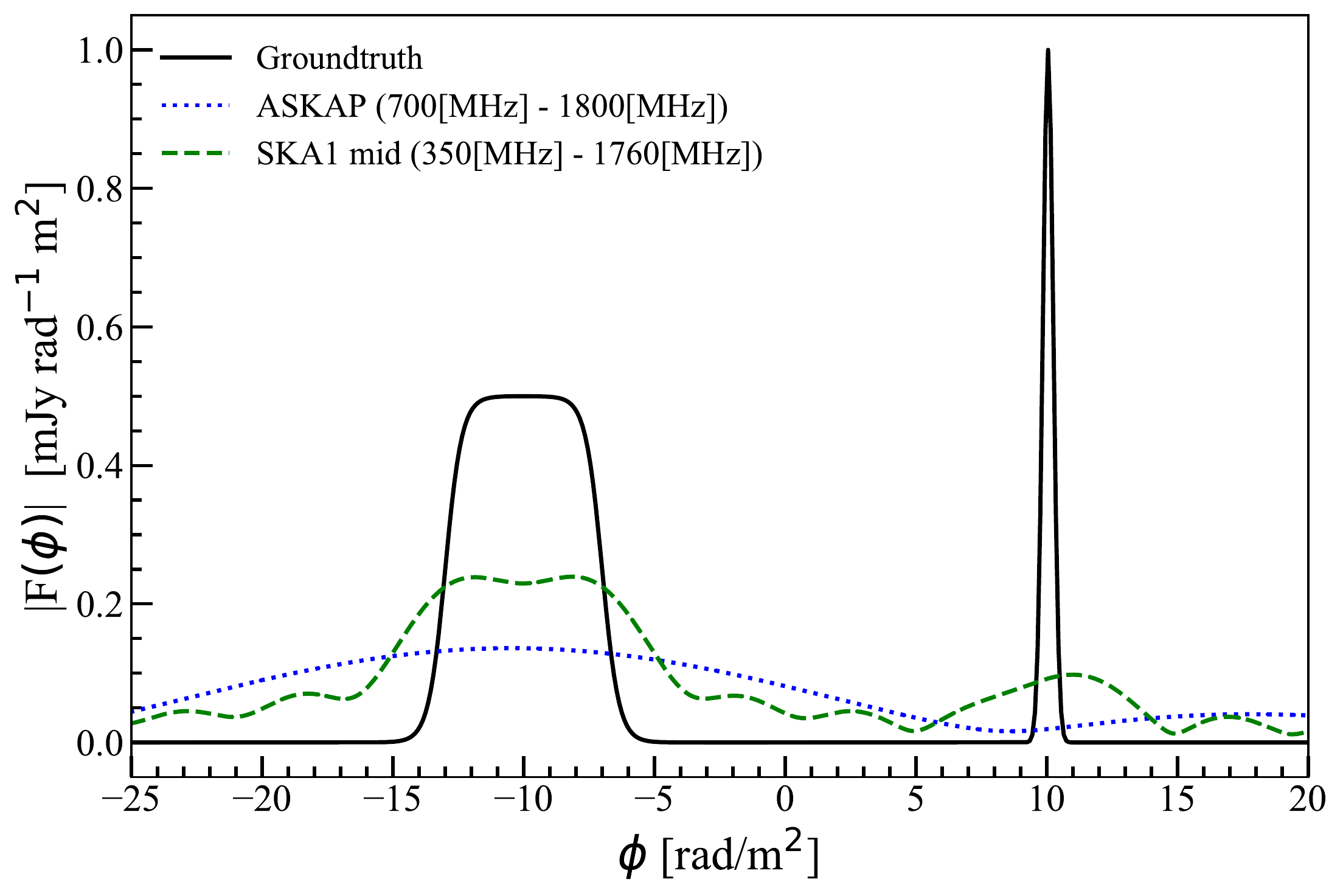}
        \caption{The amplitudes of the analytic FDF model and the RMSF-convolved FDFs for the two cases of spectral coverage as explained in Section \ref{sec:spectral_dependence}. The solid black line is the model FDF,  dotted blue line corresponds to the RMSF-convolved FDF for ASKAP frequencies,  dashed green line is the RMSF-convolved FDF for the upcoming SKA1 mid frequency range.}
        \label{fig:simple_model}
    \end{figure}

    Similarly to the previous case in Section \ref{sec:application}, the model is generated for -1000 [rad m$^{-2}$] $\le \phi <$ 1000 [rad m$^{-2}$] with a grid size of 0.1 [rad m$^{-2}$]. This FDF is then numerically Fourier transformed and data points for $\lambda^2_{\mathrm{min}} \le \lambda^2 < \lambda^2_{\mathrm{max}}$ are considered to be the polarization observation from the instrument. ASKAP has a frequency range of 700 [MHz] to 1800 [MHz], which gives $\lambda^2_{\mathrm{min}}=0.027$ [m$^{2}$] and $\lambda^2_{\mathrm{max}}=0.183$ [m$^{2}$]. SKA1 mid has a frequency range of 350 [MHz] to 1760 [MHz] for the suitable Bands 1 and 2 for Faraday tomography, which gives $\lambda^2_{\mathrm{min}}=0.029$ [m$^{2}$] and $\lambda^2_{\mathrm{max}}=0.734$ [m$^{2}$]. The FWHM of the RMSF for ASKAP and SKA1 mid coverage is 22.25 [rad m$^{-2}$] and 4.91 [rad m$^{-2}$], respectively. Figure \ref{fig:simple_model} shows the amplitude of the simple model and the RMSF-convolved counterparts for the two spectral coverages. 
    
    \begin{figure*}
        \centering
        \includegraphics[width=0.9\textwidth]{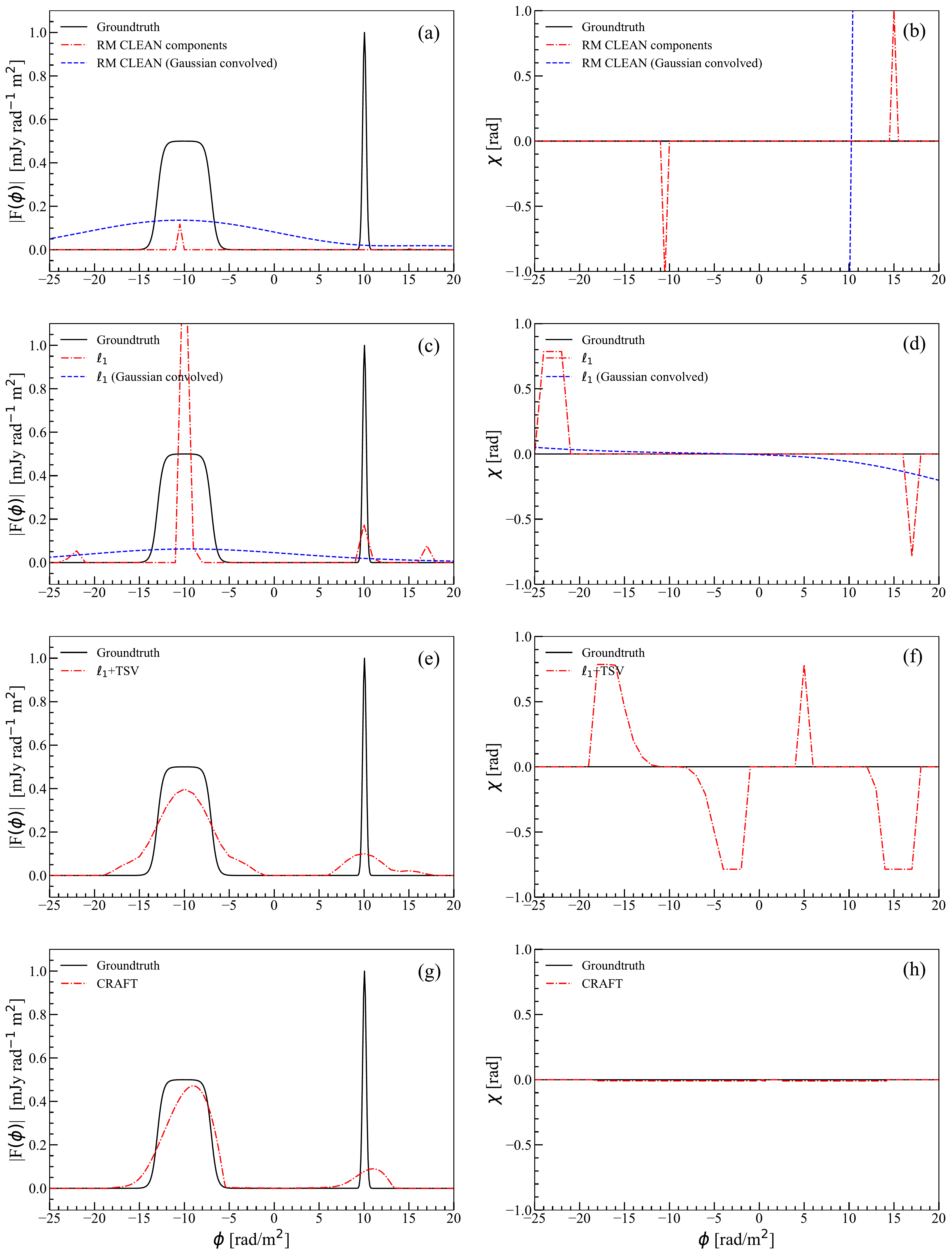}
        \caption{The figure shows the reconstruction comparison by the different Faraday tomography techniques for the model explained in Section \ref{sec:spectral_dependence} with ASKAP frequency coverage. The left panels (a), (c), (e), (g) shows the amplitudes of the FDF reconstructions by RM CLEAN, $\ell_1$ regularized reconstruction, and $\ell_1$+TSV regularized reconstruction, and CRAFT, respectively. On the other hand, the right panels (b), (d), (f), (h) are the corresponding polarization angles for the left panels. The solid black line shows the original model FDF and the red dash-dotted line show the reconstructed FDF by those techniques. The obtained FDFs by RM CLEAN and $\ell_1$ regularized reconstruction are smoothed by a Gaussian kernel with the FWHM equivalent to that of the RMSF as a common practice. The blue dashed lines correspond to the smoothed FDF amplitudes and their polarization angles. FWHM of the smoothing kernel is 22.25 [rad m$^{-2}$] for the concerned frequency range. From the Eq. (\ref{eq:recon_scale}), the smallest possible reconstruction scale is 9.44 [rad m$^{-2}$] for the observed frequency range.}
        \label{fig:recons_comparison_ASKAP}
    \end{figure*}
    
    \begin{figure*}
        \centering
        \includegraphics[width=0.9\textwidth]{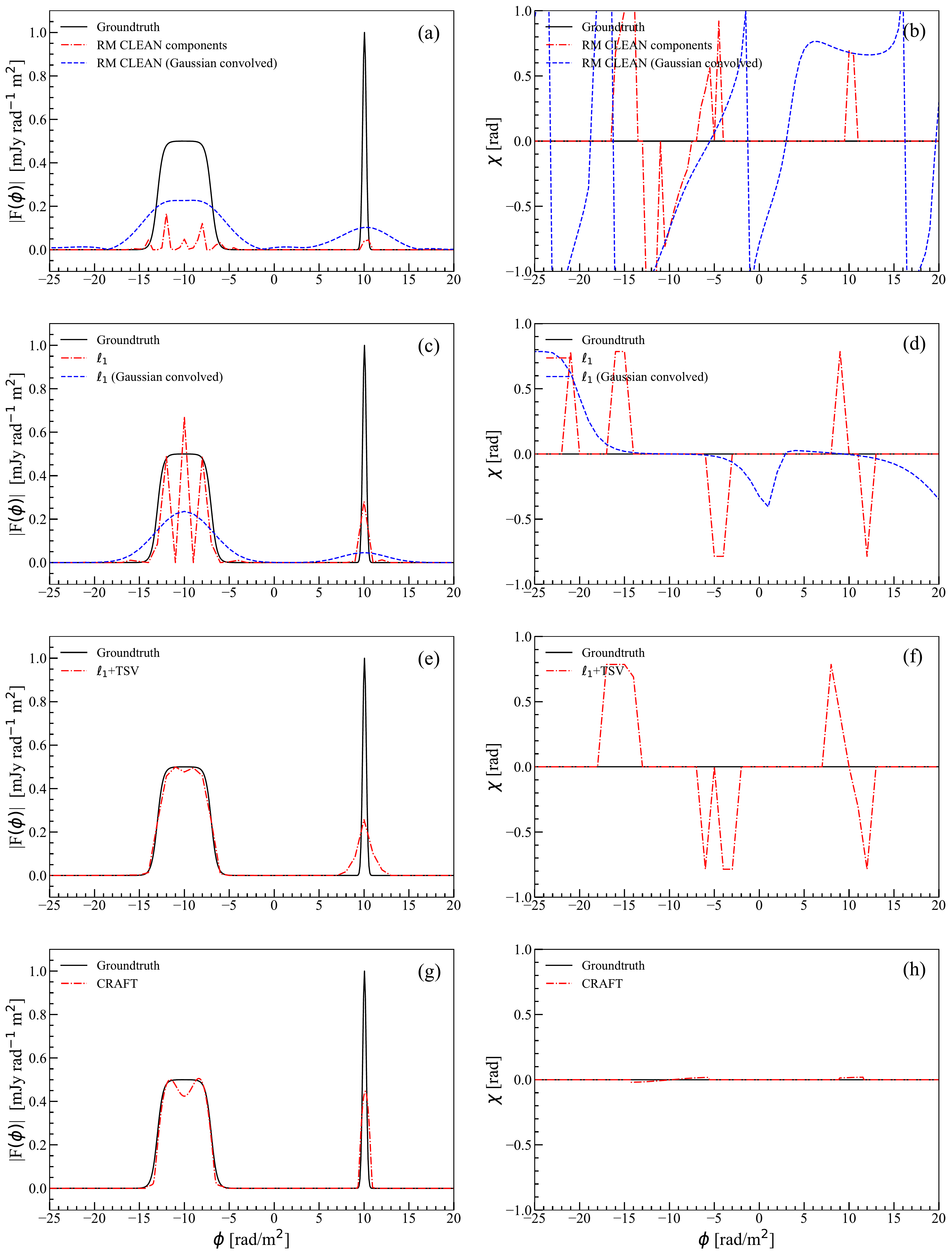}
        \caption{The figure shows the reconstruction comparison by the different Faraday tomography techniques for the model explained in Section \ref{sec:spectral_dependence} with SKA1 mid (Band 1 and 2 combined) frequency coverage. The left panels (a), (c), (e), (g) shows the amplitudes of the FDF reconstructions by RM CLEAN, $\ell_1$ regularized reconstruction, and $\ell_1$+TSV regularized reconstruction, and CRAFT, respectively. On the other hand, the right panels (b), (d), (f), (h) are the corresponding polarization angles for the left panels. The solid black line shows the original model FDF and the red dash-dotted line show the reconstructed FDF by those techniques. The obtained FDFs by RM CLEAN and $\ell_1$ regularized reconstruction are smoothed by a Gaussian kernel with the FWHM equivalent to that of the RMSF as a common practice. The blue dashed lines correspond to the smoothed FDF amplitudes and their polarization angles. FWHM of the smoothing kernel is 4.91 [rad m$^{-2}$] for the concerned frequency range. From the Eq. (\ref{eq:recon_scale}), the smallest possible reconstruction scale is 2.36 [rad m$^{-2}$] for the observed frequency range.}
        \label{fig:recons_comparison_SKAmid}
    \end{figure*}
    
    Figure \ref{fig:recons_comparison_ASKAP} shows the comparison of the methods for the above model with ASKAP frequency coverage. RM CLEAN is implemented with grid size of 0.5 [rad$ $ m$^{-2}$], gain of 0.1 and a threshold of 0.025 [mJy rad$^{-1}$ m$^2$]. $\ell_1$ sparse reconstruction is done with a grid size of 1 [rad$ $ m$^{-2}$] with $\Lambda_{\ell} = 10$, and $\ell_1$+TSV sparse reconstruction is done with a grid size of 1 [rad$ $ m$^{-2}$] with $(\Lambda_{\ell}, \Lambda_{t}) = (10, 10^3)$. CRAFT was run with $\mu = 0.01$ [mJy rad$^{-1}$ m$^2$] and a $\chi$ smoothing length scale of 40.36 [rad$ $ m$^{-2}$]. CRAFT converged in 252 iterations for $\epsilon=0.001$. It is clear that RM CLEAN and $\ell_1$ regularized reconstructions are very poor compared to the groundtruth, failing to reproduce the Faraday sources well. On the other hand, $\ell_1$+TSV regularization show some promise in reconstructing the FDF amplitudes. However, the polarization angle reconstruction is far from the expected constant of zero. CRAFT produced better confinements in $\phi$ and FDF amplitude when compared to $\ell_1$+TSV regularization. However, both the Faraday-thin and Faraday-thick components appear to be skewed, resulting in an incorrect peak Faraday depth for the components. Despite this, CRAFT produced the smallest deviation of polarization angle from the model.
    
    Figure \ref{fig:recons_comparison_SKAmid} shows the comparison for the SKA1 mid frequencies. RM CLEAN is implemented with grid size of 0.5 [rad$ $ m$^{-2}$], gain of 0.1 and a threshold of 0.025 [mJy rad$^{-1}$ m$^2$]. $\ell_1$ sparse reconstruction is done with a grid size of 1 [rad$ $ m$^{-2}$] with $\Lambda_{\ell} = 10$, and $\ell_1$+TSV sparse reconstruction with a grid size of 1 [rad$ $ m$^{-2}$] and $(\Lambda_{\ell}, \Lambda_{t}) = (10, 10^2)$. CRAFT was run with the $\mu = 0.01$ [mJy rad$^{-1}$ m$^2$] and a $\chi$ smoothing length scale of 8.92 [rad$ $ m$^{-2}$]. In this case, CRAFT converged in 92 iterations for $\epsilon=0.001$. Similar to the case for ASKAP, $\ell_1$+TSV regularized reconstruction and CRAFT provides significant improvements over RM CLEAN and $\ell_1$ regularization. Both $\ell_1$+TSV regularized reconstruction and CRAFT captures the two Faraday sources and demonstrate  agreement in the height of the Faraday-thick component. CRAFT shows its multi-scale capability in capturing the Faraday-thin component better than $\ell_1$+TSV regularization by confining tighter in $\phi$ and reproducing more of the FDF amplitude. We also note the excellent reproduction of the polarization angle in comparison to other techniques.
    
    We conducted the quantitative NRMSE analysis of the reconstruction error as also done previously. The NRMSE for each reconstruction in this section is shown in Table \ref{table:simple_nrmse}. We see a significant improvement by $\ell_1$+TSV regularized reconstruction and CRAFT compared to RM CLEAN and $\ell_1$ regularized reconstruction. CRAFT shows further improvements in NRMSE over $\ell_1$+TSV regularized reconstruction. More importantly, the significant advantage of CRAFT is the interpretability of the polarization angles. 
    
    Despite the significant improvements, we note that CRAFT does not produce perfect results, and thus more sophisticated techniques do not guarantee to fully solve the Faraday tomography problem. For example, reconstructions with ASKAP frequency coverage cannot reach the reconstruction standards of ones with SKA1 mid coverage data. Therefore, the importance of broader frequency coverage cannot be disregarded for improvements in Faraday tomography. 
    
    \begin{table}
        \centering
        \begin{tabulary}{\linewidth}{l C C} 
            \hline
            Reconstruction Method & NRMSE \newline (ASKAP frequency coverage) & NRMSE \newline (SKA1 mid frequency coverage) \\
            \hline \hline
            RM CLEAN \footnotemark[2] & {1.00}  & {1.01}\\
            $\ell_1$ regularization \footnotemark[2] & {0.62} & {0.52} \\
            $\ell_1$+TSV regularization & {0.52} & {0.35} \\
            CRAFT (This work) & \textbf{{0.48}} & \textbf{{0.26}} \\ 
            \hline
        \end{tabulary}
        \caption{A comparison of the reconstruction performances for the simple analytical model with ASKAP coverage and SKA1 mid coverage.}
        \label{table:simple_nrmse}
    \end{table}
    \footnotetext[2]{The NRMSE corresponds to the comparison of the smoothed FDFs with Gaussian kernel of FWHM = 22.25 [rad m$^{-2}$] and FWHM = 4.91 [rad m$^{-2}$] for ASKAP and SKA1 mid coverage, respectively.}

\section{Discussion} \label{sec:discussion}
    
    CRAFT provides reliable FDF reconstructions over the existing methods of Faraday tomography. In this paper, the quantitative error analysis was done by finding the NRMSE of the FDF. While this approach proves practical in assessing the overall shape and amplitudes of the reconstructed FDF, it may not be ideal for treating the reconstruction error at multiple scales (i.e., Faraday thick and thin components). If one requires a rigorous examination of the multi-scale reconstruction, we suggest NRMSE for the reconstructed complex linear polarization spectrum. However, such an NRMSE analysis can also be misleading. Even small changes in the polarization spectrum can result in substantially different FDFs, which may not be reflected in the score.
    
    A vital feature for the success of the CRAFT technique is the polarization angle smoothing. As shown in Section \ref{sec:application}, the smoothing may negatively affect some FDF features when quick polarization angle variations by different magnetic orientations are neglected. However, as seen by the demonstrations, this technique presents the closest to the intrinsic polarization angles compared to the other methods tested. Theoretical studies on FDFs and its reproducibility from observations have focused mostly on the FDF amplitude. We believe simulations with various intrinsic polarization angle models will have to be studied further to better understand the polarization angle smoothing on astrophysical outcomes.
    
    Another concern of this technique is the possible numerical inaccuracies that arise from repeated Fourier transforms. In particular, sampling in $\lambda^2$ affects the accuracy of the obtained FDF. In the tests we have demonstrated in this paper, we have considered infinitesimally narrow channels and equally spaced sampling in $\lambda^2$ space, which minimizes such inaccuracies. However, telescopes usually do not produce data with equidistant channels in the $\lambda^2$ space. Therefore, in real polarization observations, one has to use a non-uniform sampling for the Fourier transforms. An inclusion of Fourier transform algorithms with non-uniform sampling \citep[e.g.,][]{Keiner_2009} for CRAFT will be considered in the future.
    
    However, we argue that non-uniform sampling in $\lambda^2$ is not a critical issue for Faraday tomography with the present and the upcoming telescopes. The Eq. (63) of \citet{Brentjens_2005} provides the relation;
    \begin{equation}
        \left\|\phi_{\max }\right\| \approx \frac{\sqrt{3}}{\delta \lambda^{2}},
    \end{equation}
    where $\left\|\phi_{\max }\right\|$ is the maximum sensitive $\phi$ by the instrument and $\delta \lambda^2$ is the distance between two measurements in $\lambda^2$. If we decide a physically motivated maximum $|\phi|$ for the above relation, we can obtain a minimum $\delta \lambda^2$, below which we do not obtain much physical information about the FDF. In the FDF models considered here, the maximum $|\phi|$ is 1000 [rad m$^{-2}$], which corresponds to $\delta \lambda^2 \approx 0.0017$ [m$^{-2}$]. The above choice for the maximum $|\phi|$ is also applicable to other astrophysical observations of cosmic magnetic fields. 
    
    We can use the above calculated $\delta \lambda^2$ to calculate the minimum number of meaningful samples for a particular $\lambda^2$ range. For example, the number of meaningful samples for the ASKAP coverage (0.027 [m$^{2}$] $<\lambda^2<0.183$ [m$^{2}$]) will be $\sim 90$. Similarly, for the SKA1 mid (0.029 [m$^{2}$] $<\lambda^2<0.734$ [m$^{2}$]), the same number is $\sim 400$. The ASKAP Polarisation Sky Survey of the Universe's Magnetism \citep[POSSUM; ][]{Gaesnler_2010} averages the channels to have a resolution of 1 [MHz]. Thus, in theory, there are roughly 1100 sampled data points in ASKAP data to estimate a regular grid in $\lambda^2$ of at least $\sim 90$ points. Extending the argument for SKA1 mid, we need to uniformly sample $\sim 400$ data points in $\lambda^2$ from 1410 channels. The smooth/slow-varying nature of the linear polarization spectrum in Figures \ref{fig:reconstruction_spectrum} and \ref{fig:CRAFT_spectrum_wo_smoothing} confirms that even complicated FDFs with fast variations in Faraday depth space can be described by a relatively sparse sampling of the linear polarization spectrum.
    
    Though estimating a regularly sampled grid of the polarization spectrum appears easy, we note that it is more challenging when the lower frequencies are included. However, it is still feasible for the frequency ranges concerned with Faraday tomography. Additionally, a large number of channels are present in modern and future telescopes. In this regard, we propose that future surveys of Faraday tomography should strategize to irregularly sample the frequency space in a way that samples the $\lambda^2$ space uniformly. Such a uniformly resampled spectrum can then be directly used for CRAFT and other Faraday tomography techniques.
    
    A significant advantage of CRAFT is that the technique is computationally inexpensive, especially in contrast to the regularized reconstructions. For example, to reconstruct the shown models in this paper, RM CLEAN required {an order of seconds}, whereas $\ell_1$ and $\ell_1$+TSV regularized reconstructions required about 1.5 hours per FDF on modern personal computers. On the other hand, CRAFT converged in a few seconds with excellent FDF reconstructions. Efficient Faraday tomography techniques such as CRAFT will significantly facilitate the analysis of the unprecedented number of polarized emitters detected through upcoming telescopes.

\section{Conclusion} \label{sec:conclusion}

    We have presented a novel model-independent reconstruction technique for Faraday tomography called CRAFT. The method is demonstrated on a simple analytic FDF and a simulated FDF \citep{Ideguchi_2014b} of a sophisticated Milky Way model \citep{Akahori_2013}. The demonstration shows that the proposed technique can efficiently capture multi-scale features of the FDF with acceptable large-scale polarization angle reconstruction, outperforming existing popular Faraday tomography techniques.
    
    The well-accepted RM CLEAN and the more recent $\ell_1$ regularized reconstruction methods may be employed when the FDF is sparse in $\phi$, such as for Faraday thin emissions from distant radio galaxies. However, a sparsity prior in $\phi$ ($\ell_1$ penalization) may not be straightforwardly applicable for extended and mixed FDFs, such as for diffuse Milky Way emissions and the model FDFs explored here. On the other hand, $\ell_1$+TSV regularized reconstruction method intrinsically incorporates smoothness, leading to better reconstructions for extended FDFs. However, $\ell_1$+TSV regularization fails to produce Faraday-thin components simultaneously with the Faraday-thick structures. In retrospect, the currently available techniques perform better for particular scenarios while not so well at others.

    CRAFT depends on the confinement of the FDF in $\phi$, which can naturally be incorporated through other known physical constraints on cosmic magnetism. Additionally, the proposed reconstruction technique can consistently produce multi-scale features with a more physically interpretive polarization angle, which is well beyond any existing methods. The multi-scale reconstruction ability allows for a single algorithm to be used for the Faraday tomography of both galactic and extragalactic emissions. NRMSE analysis quantitatively confirms the competency of the iterative technique proposed here.
    
    This paper primarily investigated the use of model-independent techniques for Faraday tomography. The clear advantage of these techniques is that we do not have to guess the shape of the FDF (i.e., the model). The model-dependent methods, such as QU-fitting, has been successful in reconstructing simple FDFs with few components. However, as we have seen from the realistic FDF (Section \ref{sec:application}), resolved galaxy FDFs are not simple enough to be fully modeled by the commonly used analytic functions. These FDFs can have multiple peaks and varying polarization angles within the source. On the other hand, employing more model parameters in fitting can increase the chance of the local maxima problem. Model-independent Faraday tomography techniques are then the viable approach to overcome the above issues. CRAFT can particularly reconstruct both simple and complex FDFs with accurate polarization angle changes larger than the RMSF lobe width.

    In practice, the iterative reconstruction technique's performance will depend on several factors, including the complexity of the observed FDF, observing frequency coverage, noise, and the imposed constraints. Follow-up systematic studies on polarization source scales, $\lambda^2$ coverage, signal-to-noise ratios, and reconstruction constraints are necessary to identify the CRAFT algorithm's possibilities and limitations. Such systematic studies will be reported in future papers.
    
    We also note that the introduced algorithm can be easily extended to higher dimensions. In this work, we focused on the one-dimensional reconstruction along the LOS. However, by observing multiple polarization sources for a sky region, we can reconstruct a 3-dimensional distribution of the magneto-ionic media by extending the transforms to 3D. We will consider such situations in upcoming works. The reconstruction of the all-sky angular distribution from partial observations is explored in Cooray et al. in preparation.
    
    Lastly, unprecedented amounts of wide-band polarization data are now available through the Square Kilometre Array (SKA) precursors/pathfinders such as the Low Frequency Array \citep[LOFAR;][]{LOFAR}, the Murchison Widefield Array \citep[MWA;][]{MWA}, ASKAP \citep{ASKAP}, MeerKAT \citep{MeerKAT}, and the Karl G. Jansky Very Large Array \citep[VLA;][]{VLASS_2020}. Despite this, the lack of a reliable technique for Faraday tomography had noticeably plagued cosmic magnetism studies. High fidelity reconstructions, while being model-independent and computationally inexpensive, are significant strengths of this technique. Many follow-up studies and improvements will be necessary to overcome further challenges. However, the proposed CRAFT reconstruction technique shows favorable prospects to greatly facilitate cosmic magnetism studies, especially with enormous amounts of upcoming polarization data from the SKA.

\section*{Acknowledgment}

Firstly, we thank the anonymous referee for her/his careful reading of
the manuscript to provide suggestions that significantly improved this paper. This work was supported in part by JSPS Grants-in-Aid for Scientific Research (TTT: 17H01110, 19H05076, KI: 15H05890, 18K03616, and 17H01110). TTT is supported in part by the Sumitomo Foundation Fiscal 2018 Grant for Basic Science Research Projects (180923), and the Collaboration Funding of the Institute of Statistical Mathematics ``New Development of the Studies on Galaxy Evolution with a Method of Data Science''. KT is partially supported by Grand-in-Aid from the Ministry of Education, Culture, Sports, and Science and Technology (MEXT) of Japan No. 16H05999, Bilateral Joint Research Projects JSPS, and the ISM Cooperative Research Program 2020-ISMCRP-2017.

\section*{Data availability}

The data underlying this article will be shared on reasonable request to the corresponding author.



\bibliographystyle{mnras}
\bibliography{references}



\appendix
\section{FDF reconstructions without smoothing the polarization angle} \label{sec:appendix_a}

\begin{figure}
        \centering
        \includegraphics[width=0.95\linewidth]{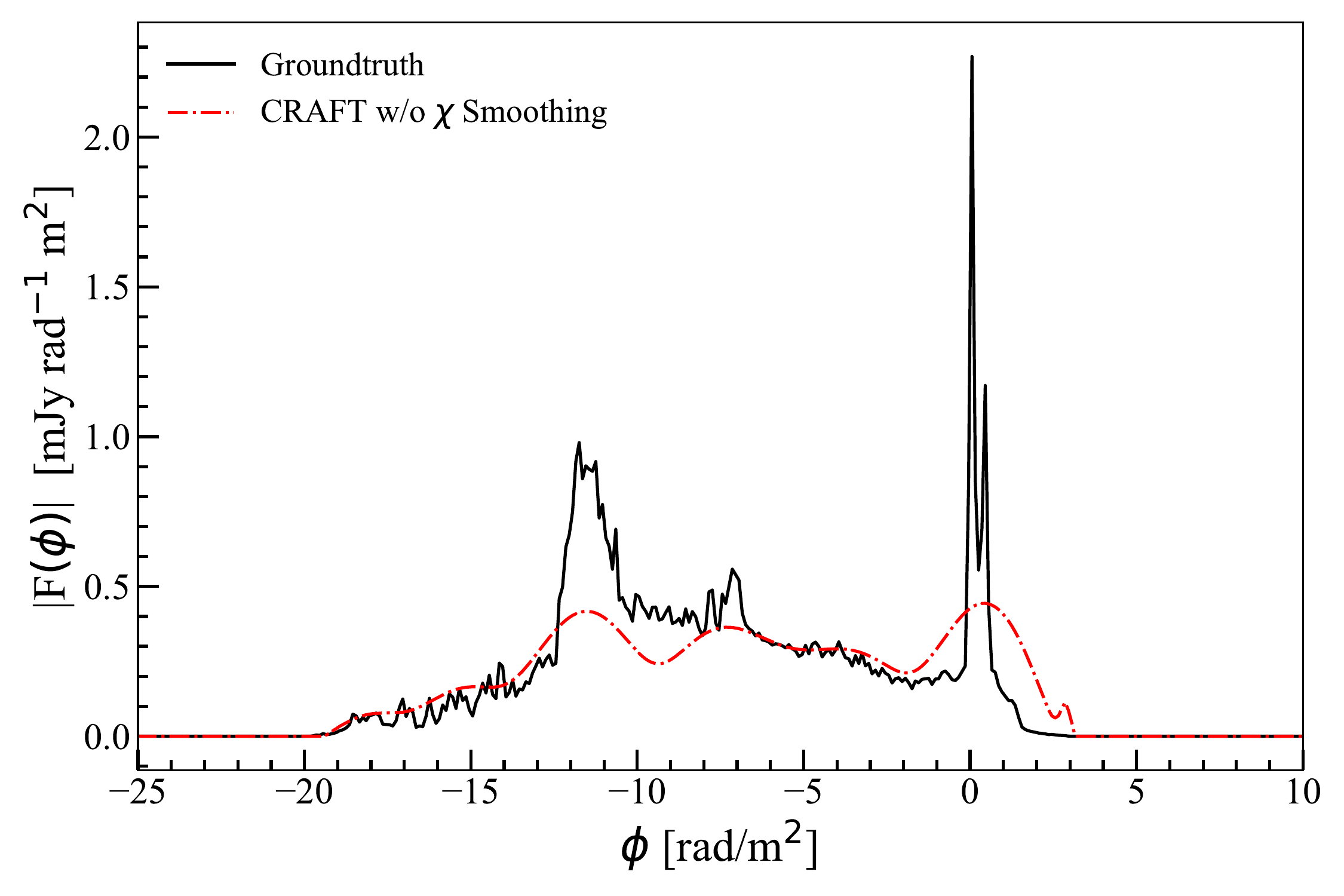}
        \caption{Amplitude of the realistic galaxy FDF reconstruction by the proposed technique without smoothing the polarization angle. Black solid line show the original model FDF and the red dash dotted line show the reconstructed FDF.}
        \label{fig:CRAFT_FDF_wo_smoothing}
    \end{figure}

\begin{figure}
        \centering
        \includegraphics[width=0.95\linewidth]{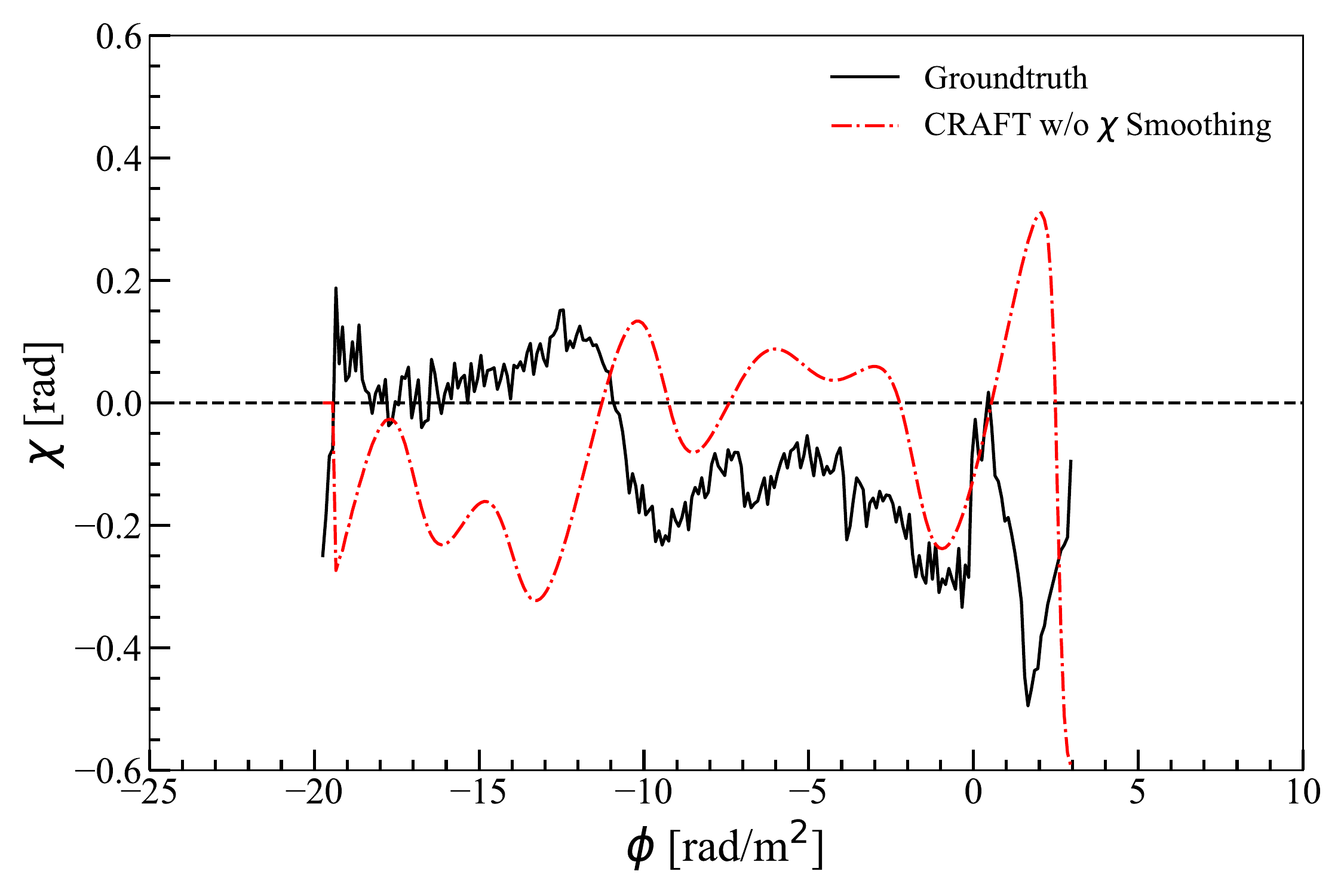}
        \caption{Polarization angle of the CRAFT reconstructed realistic galaxy FDF but without smoothing the polarization angle. Black solid line show the synthetic simulation and the red dash dotted line show the reconstructed. }
        \label{fig:CRAFT_chi_wo_smoothing}
    \end{figure}

\begin{figure} 
        \centering
        \includegraphics[width=0.95\linewidth]{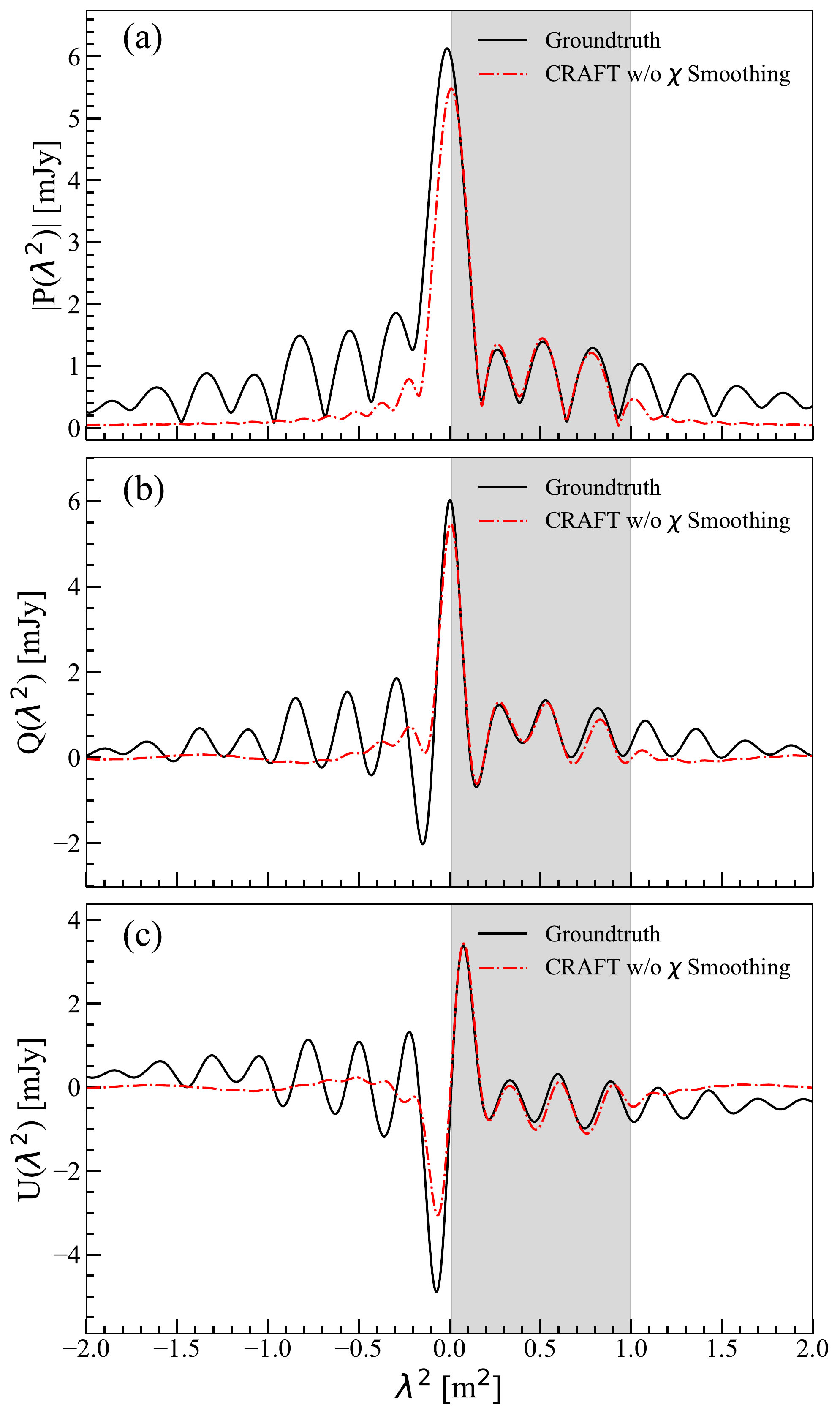}
        \caption{CRAFT reconstructed linear polarization spectrum of the realistic galaxy FDF. Panel (a) shown the amplitude of the complex linear polarization spectrum, (b) is the  Stokes Q, and (c) is the Stokes U. The black dotted line is the of the noiseless synthetic model FDF and the red dash dotted line show its reconstruction by the CRAFT technique without smoothing the polarization angle. Observed $P(\lambda^2)$ are approximately between 0.01 [m$^2$] and 1.00 [m$^2$], shown by the shaded region.}
        \label{fig:CRAFT_spectrum_wo_smoothing}
    \end{figure}

    In Figures \ref{fig:CRAFT_FDF_wo_smoothing}, \ref{fig:CRAFT_chi_wo_smoothing}, and \ref{fig:CRAFT_spectrum_wo_smoothing}, we show the CRAFT reconstruction of the realistic galaxy FDF without smoothing the polarization angle. The NRMSE for this reconstruction is 0.60.%
    
    We see in Figure \ref{fig:CRAFT_FDF_wo_smoothing}, that despite capturing the multi-scale components in the reconstructed FDF, amplitudes (peaks) are not well reconstructed. The difficulty of reconstruction is apparent for the polarization angle shown in Figure \ref{fig:CRAFT_chi_wo_smoothing}. We see unsatisfactory reconstruction far from the truth and small scale fluctuations. The shortcomings are more evident when we look at the linear polarization spectrum in Figure \ref{fig:CRAFT_spectrum_wo_smoothing}. It has failed to reconstruct the negative $\lambda^2$ side of the spectrum due to the insufficient constraints.

    
\section{Comparisons with Smoothed Synthetic Simulation} \label{sec:appendix_b}
    
\begin{figure}
        \centering
        \includegraphics[width=0.95\linewidth]{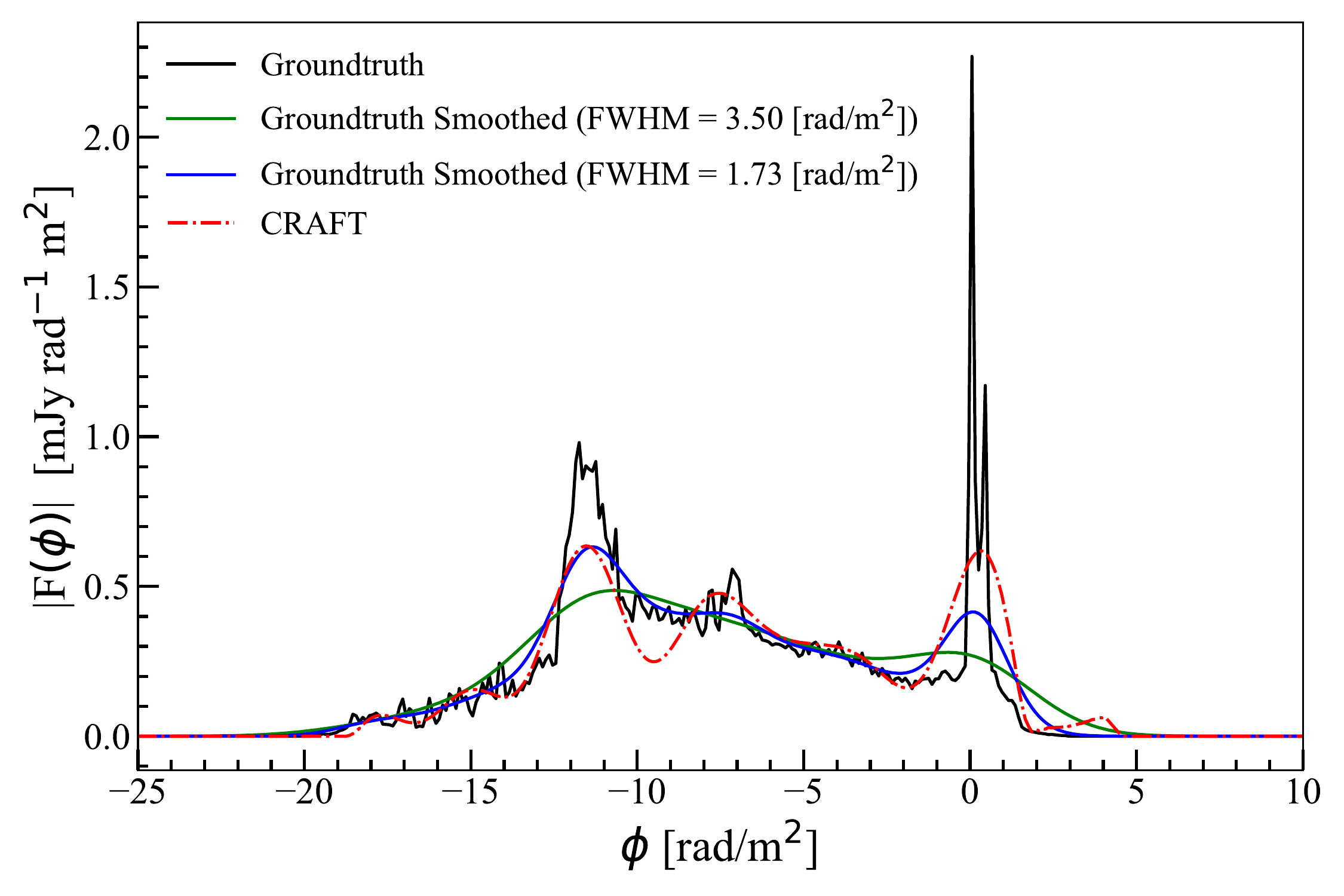}
        \caption{A comparison of the CRAFT FDF reconstruction with the smoothed model FDFs. Black solid line show the original model FDF, the green solid line shows the model FDF smoothed by a Gaussian kernel of FWHM equivalent to the RMSF FWHM (3.50 [rad m$^{-2}$]), blue solid line shows the Gaussian smoothed model FDF at the minimum reconstruction scale (1.73 [rad m$^{-2}$]), and the red dash dotted line show the reconstructed FDF by CRAFT.}
        \label{fig:model_smoothed_CRAFT}
    \end{figure}

    We compare the reconstruction by CRAFT with the smoothed model FDFs. Figure \ref{fig:model_smoothed_CRAFT} shows the amplitudes of the original synthetic model FDF, two smoothed model FDFs at different scales in $\phi$, and the CRAFT result. As a common practice, reconstructions with RM CLEAN or $\ell_1$ regularized reconstructions require smoothing with a Gaussian kernel of FWHM = RMSF FWHM for better results. In comparison, the CRAFT result is significantly better than the model FDF that is Gaussian smoothed at FWHM = RMSF FWHM = 3.50 [rad m$^{-2}$] scale, which has an NRMSE of 0.53. On the other hand, the CRAFT result should resemble the model FDF smoothed at the minimum reconstruction scale (1.73 [rad m$^{-2}$]) from the observed data as defined in Eq. (\ref{eq:recon_scale}). The CRAFT reconstruction shows general agreement in the extended component, and the peak between $\phi \approx -13$ [rad m$^{-2}$] and $\phi \approx -11$ [rad m$^{-2}$] is also satisfied. Furthermore, CRAFT has reproduced the Faraday-thin peak at $\phi \approx 0-1$ [rad m$^{-2}$] better than the smoothed model. We attribute the few inconsistencies of the CRAFT result with the smoothed model to the incomplete polarization angle reconstruction. However, CRAFT significantly improves our ability to reconstruct beyond the RMSF FWHM resolution by reconstructing the negative $\lambda^2$ side, at least up to $-\lambda^2_{\textrm{obs, max}}$.
    
\bsp    
\label{lastpage}
\end{document}